\documentclass[12pt,fleqn]{article}
\usepackage{latexsym,amsfonts,amssymb}
\newcommand{\be}[1]{\begin{equation}\label{#1}}
\newcommand{\ee}{\end{equation}}
\newcommand{\bs}[1]{\begin{equation}\label{#1}\arraycolsep=0em\begin{array}{l}}
\newcommand{\es}{\end{array}\end{equation}}
\newcommand{\bss}{\[\arraycolsep=0em\begin{array}{l}}
\newcommand{\ess}{\end{array}\]}
\newcommand{\ba}[1]{\begin{array}{#1}}
\newcommand{\ea}{\end{array}}
\newcommand{\bc}{\begin{center} }
\newcommand{\ec}{\end{center} }
\newcommand{\bt}[1]{\begin{tabular}{#1}}
\newcommand{\et}{\end{tabular} }

\renewcommand{\ge}{\geqslant}

\newcommand{\const}{\mathop{\rm const}\nolimits}

\newcommand{\tfrac}[2]{\textstyle\frac{#1}{#2}}
\newcommand{\p}{\partial}
\newcommand{\R}{{\mathbb R}}

\begin{document}
\large

\begin{flushleft}
\bf Roman O. Popovych${}^\dag$ and  Irina A. Yehorchenko${}^\ddag$
\end{flushleft}

\begin{flushleft}
Institute of Mathematics, National Academy of Science of Ukraine,
\\ 3 Tereshchenkivs'ka Street, 01601, Kyiv-4, Ukraine\\
$\dag$~E-mail: rop@imath.kiev.ua\\
$\ddag$~E-mail: iyegorch@imath.kiev.ua
\end{flushleft}

\noindent
{\Large\bf Group Classification of Generalised Eikonal Equations}

\vspace{2ex}

\begin{abstract}
A new approach to the problem of group classification is applied to the class of
first-order non-linear equations of the form $u_a u_a=F(t,u,u_t)$. It allowed
complete solution of the group classification problem for a class of equations
for functions depending on
multiple independent variables, where highest derivatives enter nonlinearly.
Equivalence groups of the class under consideration and algebraic properties of
the symmetry algebra are studied.

The class of equations considered presents generalisation of the eikonal and
Hamilton—Jacobi equations. The paper contains the list of all non-equivalent
equations from this class with symmetry extensions, and proofs of such non-
equivalence.

New first order non-linear equations possessing wide symmetry groups
were constructed.
\end{abstract}

\noindent
{\bf 1. Introduction.}
The problem of group classification of differential equations is determination
of all particular equations from the class possessing wider symmetry when the
class in general. Mathematical models, and in particular, differential equations
of
mathematical physics, often contain parameters
(numerical or functional), that are determined experimentally.
For this reason the parameters are not fixed exactly.
Theoretical selection of mathematical models can be done on the
basis of the symmetry principle, that is to select
among given values of parameters ones that lead to
a mathematical model with given properties, or with the widest
possible symmetry group~\cite{ovsiannikov.eng}.

We considered a class of mathematical models given by
{\it a rather general set of non-linear first-order equations
that are invariant under the Euclid group}~-- the group
of space rotations and translations. We did not fix any
symmetry with respect to time and dependent variables.
The set of mathematical models considered
incorporates models with relativistic and Galilei
symmetries that are well-known and widely used in
physics, geometric optics and other applications.

The basics for the group classification were constructed by S.~Lie himself
\cite{lie}
in his classification of integrable ordinary differential equations.
The start of systematic study of similar problems for partial differential
equations
is linked to papers by L.Ovsyannikov (see \cite{ovsiannikov.eng} and references
wherein) and by his collaborators.
In particular, Ovsyannikov introduced such important concepts as arbitrary
element, kernel of principal groups,
equivalence group and a number of others.

At present there are quite a few papers on group classification of important
classes of models in mechanics, physics, biology and other sciences
(see e.g. \cite{akhatov&gazizov&ibragimov.dan.1987.eng}--
\cite{zb2000.with.cherniha},
and these present only a small part of existing papers!).

Until recently the principal method for solving of group
classification problems for partial differential equations was direct
integration of determining equations for a Lie symmetry
operator with sorting out all possible cases of
integration. Application only of the above method
substantially narrowed the scope of solvable group
classification problems, as determining equations may
be extremely difficult to solve, if it is possible at all.
Most of the group classification problems solved in
the literature are problems for classes of equations
with arbitrary functions on only one variable each.
Arbitrary functions with multiple variables result in determining
equations being partial differential equations, rather
than ordinary. It is obvious that this complicates integration considerably.

Recently new approaches appeared having allowed solution
of group classification problems that could not be solved by means of the
methods known before.
Let us mention in more detail a few papers whose ideas are the closest to this
paper.

I.S.~Akhatov, R.K.~Gazizov and N.K.~Ibragimov
\cite{akhatov&gazizov&ibragimov.eng},
developing the approach by L.Ovsyannikov, systematically employed the
equivalence group for simplification of classifying conditions.

R.~Zhdanov and V.~Lahno \cite{zhdanov&lahno.jphys.a.1999}
suggested a new approach to group classification of PDEs
that presents a generalisation of S.~Lie's approach to
group classification of ODEs~\cite{lie} (see
also~\cite{zhdanov&roman.rep.math.phys.2000}). This
approach is based on classification of low-dimensional abstract algebrae and
works best for equations with two independent variables.
It allows avoiding direct solving of cumbersome determining equations and
finding complete solutions
for non-linear group classification problems for parameter functions depending
on many variables.
Determining equations to be solved are partial differential
equations with respect to the parameter functions.
Further development for this approach
presented in this paper is applied for equations with arbitrary number of
independent variables.
R.~Popovych and R.~Cherniha~\cite{zb2000.with.cherniha} employed for
classification
of systems of coupled non-linear Laplace equations
a combined method in which the number of cases that are
considered during integration of determining equations,
is essentially decreased at the expense of the study of
possible structure of the symmetry algebra. There are a number of other papers
suggesting new approaches to group classification problems.

In the present paper the combined method is used for group
classification of generalised eikonal and Hamilton-Jacobi
equations. The results of the paper show that this method
that includes investigation of determining equations,
of equivalence transformation groups and of the symmetry
algebra structures is an efficient tool for finding
equations from a certain class with symmetry extensions.
However, the method is not strictly algorithmic,
and the steps to be used would be different for different
classes of equations.

We succeeded to give a full description of subclasses for
the class considered where symmetry extension
arises compared to the general symmetry group for the class
in total. We expect that the method suggested
would allow solving other group classification problems,
especially for higher order equations and for systems
of PDEs, and extending the variety of non-linear equations
with wide symmetry that could be used for construction
of mathematical models in physics and other natural sciences.

\vspace{1ex}

\noindent
{\bf 2. Formulation of the Problem.} Let us consider the equation of the form
\be{gehje}
u_au_a=F(t,u,u_t)
\ee
for a real function $u=u(t,x)$ of $n+1$ independent variables
$t=x_0$ and $x=(x_1, x_2, \ldots, x_n)$, $n\ge 2$.
Here and below a lower index of a function will designate
differentiation with respect to the corresponding variable.
The indices $a$ and $b$ take values from 1 to $n$.
The summation over repeated indices is implied unless stated otherwise.
The class of equations (1) is the general class of non-linear
first-order equations invariant under the Euclid group~--
the group of space rotations and translations. It is
remarkable also as it includes all the following well-known equations:
\[\arraycolsep=0em\begin{array}{ll}
F=2mu_t&\mbox{ is  the Hamilton-Jacobi equation for a free particle;}\\
F=u_t^2-1&\mbox{ is  the relativistic Hamilton equation;}\\
F=u_t^2&\mbox{ is  the eikonal equation.}\\
\end{array}\]
The symmetry properties of these equations and relations between them were
investigated in
\cite{boyer&penafiel,fushchich&shtelen.lett.nuovo.cim.1982,fshs.eng}.
Equations of the form $u_t^2=f(u) u_au_a$ that are equivalent to the eikonal
equation, were used in
\cite{fs.danu.90.7}
for studying of solutions of non-linear wave equations.

We perform group classification for equations of the form~(\ref{gehje}) by an
arbitrary element --- a smooth function $F=F(t,u,u_t)\not\equiv0.$
(Here and below smoothness means continuous differentiability.)

\vspace{2ex}

\noindent
{\bf 3. Remarks on Formulation of Classification Problems and
Sche\-mes for their Solving.} We would like to explain our choice
of the class we consider for group classification. This choice
was governed by the following rules:

1. A class has to generalise physically interesting equations.

2. A problem has to be symmetric by itself: a class under
consideration should have a wide equivalence group.

3. A class has to preserve at least some of important
symmetry properties of equations being generalized.

The second rule was the basis of the approach suggested
in~\cite{zhdanov&lahno.jphys.a.1999,zhdanov&roman.rep.math.phys.2000}.
Both second and third rules
were the reason why we included dependence on the variable
$t$ for the arbitrary function $F$. That preserved the
possibility of symmetries interchanging $t$ and $u$ that
the listed well-known equations have. Introduction of $t$
substantially increases the equivalence group, and despite
most resulting classes with symmetry extensions could be
represented by $F=F(u,u_t)$, or $F$ does not depend on $t$,
the equivalence transformations employed in the classification
process belonged to the wide equivalence group corresponding
to the class with $F=F(t,u,u_t)$.

The general outline of our classification could be presented as follows.

1. Find determining equations and the kernel of main groups.

\vspace{0.5ex}

2. Find the equivalence group.

\vspace{0.5ex}

3. Integrate a part of the system of determining equations and
obtain a possible form for invariance algebra operators.

\vspace{0.5ex}

4. Classify possible symmetry algebrae that are lowest-order
extensions of the main algebra and that possess a certain
structure identified as a separate case at the step 3.
Distinguish the corresponding non-equivalent cases.

\vspace{0.5ex}

5. Impose conditions for the invariance algebrae having
larger dimension or different structure than those
considered at the step 4, find new cases. The requirement
of a different structure than was considered before
allowed to study remaining cases by means of the standard approach.

\vspace{1ex}

\noindent
{\bf 4. Kernel of Main Groups and Equivalence Group.}
Let an infinitesimal operator
\[
Q=\xi^0(t,x,u)\p_t+\xi^a(t,x,u)\p_a+\eta(t,x,u)\p_u
\]
generate a one-parameter symmetry group of local transformation for the
equation~(\ref{gehje}).
Then from the infinitesimal invariance criterion
\cite{ovsiannikov.eng,olver.eng} after transition to the manifold given by the
equation ~(\ref{gehje})
in the prolonged space, and splitting by non-related variables we derive the
following determining
equations for the coefficients of the operator $Q$:
\be{det.eqs.for.gehje.1}
\xi^a_b+\xi^b_a=0, \qquad \xi^a_a=\xi^b_b, \qquad a\not=b
\ee
(here there is no summation by $a$ and $b$),
\be{det.eqs.for.gehje.2}
(\xi^a_t+\xi^a_uu_t)F_{u_t}-2\xi^a_uF+2\eta_a-2u_t\xi^0_a=0,
\ee
\be{det.eqs.for.gehje.3}
\xi^0F_t+\eta F_u+(\eta_t+(\eta_u-\xi^0_t)u_t-\xi^0_uu_t^2)F_{u_t}=2(\eta_u-
\xi^1_1-\xi^0_uu_t)F.
\ee

If the function $F$ is not fixed, then splitting in~(\ref{det.eqs.for.gehje.2})
and~(\ref{det.eqs.for.gehje.3})
by the ``variables'' $F$, $F_t$, $F_u$, $F_{u_t}$, $u_t$,
we obtain that $\eta=\xi^0=0$, $\xi^1_1=0$, $\xi^a_t=\xi^a_u=0$, whence taking
into
account~(\ref{det.eqs.for.gehje.1}) we come to the following statement.

\vspace{1ex}

\noindent
{\bf Statement.} The kernel of the main groups for the equation~(\ref{gehje}) is
the Euclid group $E(n)$,
the algebra of which is
$
A^{\rm ker}=e(n)=\langle\p_a, \; J_{ab}=x_a\p_b-x_b\p_a\rangle.
$

\vspace{1ex}

The equivalence group for the equation~(\ref{gehje}) coincides with the group
generated by the set of
one-parameter groups of local symmetries of the system
\be{system.equiv.gehje}
u_au_a=F, \qquad w=u_t, \qquad F_a=0,
\ee
whose infinitesimal operators have the form
\bss
\widehat Q=\hat \xi^0(t,x,u)\p_t+\hat \xi^a(t,x,u)\p_a+\hat
\eta(t,x,u)\p_u\\[1ex]
\phantom{\widehat Q={}}+\hat \theta(t,x,u,w)\p_w+\hat \chi(t,x,u,w,F)\p_F.
\ess
From the infinitesimal invariance criterion for the
system~(\ref{system.equiv.gehje})
after splitting with respect to non-related variables we obtain determining
equations
for coefficients of the operator $\widehat Q$, from which it follows that
an infinitesimal operator of any one-parameter equivalence group for the
equation~(\ref{gehje}) is
a linear combination of operators
\be{basis.operators.of.gehje.equivalence.algebra}
\p_a, \quad J_{ab}, \quad x_a\p_a-2F\p_F, \quad
\bar\xi\p_t+\bar\eta\p_u+2(\bar\eta_u-\bar\xi_uu_t)F\p_F,
\ee
where $\bar\xi$ and $\bar\eta$ are arbitrary smooth functions of
variables $t$ and $u$.
Thus, equivalence transformations, that act non-trivially on the parameter $F$,
have the form
\bs{general.equivalence.transformation.for.gehje}
\tilde t=\zeta(t,u), \quad \tilde u=\varphi(t,u), \quad \tilde x=\delta x,
\\[1ex]\displaystyle
\tilde F=\delta^{-2}(\varphi_u-\zeta_u\tilde u_{\tilde t})^2F=
\delta^{-2}\left(\frac{\zeta_t\varphi_u-
\zeta_u\varphi_t}{\zeta_t+\zeta_uu_t}\right)^2F,
\es
 where $\delta$ is a non-zero constant,
$\zeta$ and  $\varphi$ are arbitrary smooth functions of
variables $t$ and $u$, for which
$\zeta_t\varphi_u-\zeta_u\varphi_t\not=0$.

If we limit the class of equations~(\ref{gehje}), having imposed an additional
condition $F_t=0$
(this subclass is separated in the process of classification),
then for calculation of the corresponding equivalence group it is necessary to
add this condition
to the system~(\ref{system.equiv.gehje}).
As a result, the equivalence group is narrowed:
an infinitesimal operator of any one-parameter equivalence group for the
equation~(\ref{gehje})
with $F=F(u,u_t)$ is a linear combination of the operator $t\p_t$ and
of operators~(\ref{basis.operators.of.gehje.equivalence.algebra}),
where the functions $\bar\xi$ and $\bar\eta$ now depend only on the variable
$u$,
and therefore equivalence transformations that act non-trivially on the
parameter $F$ have the form
\bs{partial.equivalence.transformation.for.gehje}
\tilde t=\hat\delta t+\zeta(u), \quad \tilde u=\varphi(u), \quad \tilde x=\delta
x,  \\[1ex]\displaystyle
\tilde F=\delta^{-2}(\varphi_u-\zeta_u\tilde u_{\tilde t})^2F=
\delta^{-2}\left(\frac{\hat\delta\varphi_u}{\hat\delta+\zeta_uu_t}\right)^2F,
\es
where $\delta,$ $\hat\delta$ are non-zero constants,
$\zeta$ and $\varphi$~ are arbitrary smooth functions of the variable $u$, and
$\varphi_u\not=0$.

Further limitation of the class of equations~(\ref{gehje}) may lead to
appearance of equivalence
transformations of the form
~(\ref{general.equivalence.transformation.for.gehje}), that are
different from~(\ref{partial.equivalence.transformation.for.gehje}) (see the
proof).

\vspace{1ex}

\noindent
{\bf 5. Result of Classification.}
All possible cases of extension of the maximal invariance algebra in the Lie
sense for the
equation~(\ref{gehje}) up to equivalence
transformations~(\ref{general.equivalence.transformation.for.gehje})
are exhausted by the cases listed in the Table~1.

In the Table~1 $F=F(t,u,u_t)$, $f=f(u,u_t)$, $h=h(u_t)$ are arbitrary smooth
functions of their arguments,
$\delta$ is constant, $\varepsilon_1,\varepsilon_2=\pm 1$,
$(\varepsilon_1,\varepsilon_2)\not=(-1,-1)$,
$J_{ab}=x_a\p_b-x_b\p_a$, $D=t\p_t+u\p_u+x_a\p_a$.
In the case~7 $g_{\mu\nu}$ is metric tensor of Minkowsky space $\R^{1,n}$,
i.e. $g_{00}=-g_{11}=-g_{22}=-g_{33}=1$, $g_{\mu\nu}=0$, $\mu\not=\nu$;
$c^\mu$, $b^{\mu\varkappa}$, $d$, $a^\varkappa$, $\eta$ are arbitrary smooth
functions of the
variable $u$ (for the set of such operators to form an algebra,
it is necessary to require that these functions were infinitely differentiable
or real analytical);
the indices $\mu$, $\nu$ and $\varkappa$ take values from 0 to 3.

Let us note that the known Hamilton-Jacobi and eikonal equations, and the
relativistic Hamilton
equation are selected in the family of equations~(\ref{gehje}) as
representatives of the classes of
equivalent equations having the widest symmetry.

\vspace{1ex}

{\small \hfill Table 1}

\bc\renewcommand{\arraystretch}{1.4}\normalsize
\begin{tabular}{|r|p{3.4cm}|p{11cm}|}
\hline
&\hfil$F$\hfil&\hfil Basis Symmetry Operators \hfil \\ \hline
$\!0\!$&$F(t,u,u_t)$&$\p_a$, $J_{ab}$\\ \hline
$\!1\!$&$e^{\delta t}\!f(u,u_t)$, $\!\delta\!\in\!\{0;1\}\!\!\!$&$\p_a$,
$J_{ab}$, $2\p_t-\delta x_a\p_a$\\ \hline
$\!2\!$&$e^u h(u_t)$&$\p_a$, $J_{ab}$, $\p_t$, $2\p_u-x_a\p_a$\\ \hline
$\!3\!$&$|u|^{2-\delta} h(u_t)$, $\delta\not=2$&$\p_a$, $J_{ab}$, $\p_t$,
$2t\p_t+2u\p_u+\delta x_a\p_a$\\ \hline
$\!4\!$&$h(u_t)$&$\p_a$, $J_{ab}$, $\p_t$, $\p_u$, $D$\\ \hline
$\!5\!$&$e^{u_t}$&$\p_a$, $J_{ab}$, $\p_t$, $\p_u$, $D$, $x_a\p_a-2t\p_u$\\
\hline
$\!6\!$&$|u_t|^\beta$, $\beta\not=0,1,2$&$\p_a$, $J_{ab}$, $\p_t$, $\p_u$, $D$,
    $(\beta-2)x_a\p_a-2u\p_u$\\ \hline
$\!7\!$&$u_t^2$&
$2g_{\mu\nu}c^\mu(u)x_\nu x_\varkappa\p_\varkappa
-c^\varkappa(u)g_{\mu\nu}x_\mu x_\nu\p_\varkappa
+g_{\mu\nu}b^{\mu\varkappa}(u)x_\nu\p_\varkappa+$\hfil${ }$
$+d(u)x_\varkappa\p_\varkappa+a^\varkappa(u)\p_\varkappa+\eta(u)\p_u$\\ \hline
$\!8\!$&$\varepsilon_2 u_t^2+\varepsilon_1$&
$\p_a$, $J_{ab}$, $\p_t$, $\p_u$, $D$,
$J_{ua}=u\p_a+\varepsilon_1x_a\p_u$, $J_{ta}=t\p_a+\varepsilon_2x_a\p_t$,
$J_{ut}=u\p_t-\varepsilon_1\varepsilon_2t\p_u$,
$K_a=2x_aD-s^2\p_a$, $K_u=2uD+\varepsilon_1s^2\p_u$,
$K_t=2tD+\varepsilon_2s^2\p_t$,
 where $s^2=x_ax_a-\varepsilon_1u^2-\varepsilon_2t^2$\\ \hline
$\!9\!$&$\varepsilon_2 e^uu_t^2+\varepsilon_1$&
$\p_a$, $J_{ab}$, $\p_t$, $t\p_t+2\p_u$,
$(t^2-4\varepsilon_1\varepsilon_2e^u)\p_t+4t\p_u$ \\ \hline
$\!10\!$&$\cos^{-2}\!u\,u_t^2+1$&
$\p_a$, $J_{ab}$, $\p_t$, $\cos t\tan u\,\p_t-\sin t\,\p_u$, $\sin t\tan
u\,\p_t+\cos t\,\p_u$ \\ \hline
$\!11\!$&$\pm(\cos^{-2}\!u\,u_t^2-1)$&
$\p_a$, $J_{ab}$, $\p_t$, $\cosh t\tan u\,\p_t+\sinh t\,\p_u$, $\sinh t\tan
u\,\p_t+\cosh t\,\p_u$ \\ \hline
$\!12\!$&$\cosh^{-2}\!u\,u_t^2+1$&
$\p_a$, $J_{ab}$, $\p_t$, $\cosh t\tanh u\,\p_t-\sinh t\,\p_u$, $\sinh t\tanh
u\,\p_t-\cosh t\,\p_u$ \\ \hline
\end{tabular}\ec

\vspace{1ex}

\noindent
{\bf 6. Proof.} The complete proof is very cumbersome. We omit technical
calculations, and some
interim results are given as lemmas, and without proofs.

Let us denote the maximal invariance algebra in the Lie sense for
the equation~(\ref{gehje}) as $A^{\rm max}$.
In investigation of determining equations~(\ref{det.eqs.for.gehje.1})--
(\ref{det.eqs.for.gehje.3})
two essentially different cases arise:
$F_{u_tu_tu_t}\not=0$ and $F_{u_tu_tu_t}=0$.

\vspace{1ex}

If {\mathversion{bold}$F_{u_tu_tu_t}\not=0$}, then it follows
from~(\ref{det.eqs.for.gehje.2})
and~(\ref{det.eqs.for.gehje.3}), that $\eta_a=\xi^0_a=\xi^a_t=\xi^a_u=0$,
$\xi^1_{1a}=0$, thus $A^{\rm max}=A^{\rm ker}+A^{\rm ext}$, and
$A^{\rm ext}\subset\langle \delta x_a\p_a+ \xi^0(t,u)\p_t+\eta(t,u)\p_u\rangle$,
where $ \delta$ is an arbitrary constant,
$\xi$ and $\eta$ are arbitrary smooth functions of the variables~$t$ and~$u$.
$A^{\rm ker}$ ia an ideal, and $A^{\rm ext}$ is a subalgebra of the algebra
$A^{\rm max}$,
and $\forall Q\!\in\!A^{\rm ext} \: (Q\not=0)\mbox{:}\: (\xi^0,\eta)\not=(0,0).$
The dimension of $A^{\rm ext}$ determines dimension of an extension of the
algebra $A^{\rm max}$.

If $\dim A^{\rm ext}>0$, then we may choose any non-zero operator~$Q$
from~$A^{\rm ext}$.
By equivalence
transformations~(\ref{general.equivalence.transformation.for.gehje})
it is always possible to reduce it to the form $Q=2\p_t-\delta x_a\p_a$, where
$\delta\!\in\!\{0;1\}$, after that,
having solved equations~(\ref{det.eqs.for.gehje.3}) with $\xi^0=2$, $\xi^a=-
\delta x_a$, $\eta=0$
with respect to $F$, we obtain the first case of extension of the
algebra~$A^{\rm max}$.

Let $\dim A^{\rm ext}>1$. Then $A^{\rm ext}$ has to contain an operator $Q$ with
$\delta=0$.
By equivalence
transformations~(\ref{general.equivalence.transformation.for.gehje})
we can reduce the operator $Q$ to the operator $\p_t$. In the following we will
assume that $\p_t\in A^{\rm ext}$,
whence $F_t=0$, i.e. $F=F(u,u_t)$, and we will say that there is an additional
extension of symmetry,
if for such $F$ the relation $\dim A^{\rm max}>\dim A^{\rm ker}+1$ is satisfied.

Let us assume that~$\dim A^{\rm ext}=2,$ and
\[A^{\rm ext}=\langle Q^1=\p_t,\; Q^2=\delta
x_a\p_a+\xi^0(t,u)\p_t+\eta(t,u)\p_u\rangle\
\quad \mbox{and} \quad \eta\not=0.\]

If the algebra $A^{\rm ext}$ is commutative, then $\xi^0_t=0$, $\eta_t=0$, and
by means of the
transformation~(\ref{partial.equivalence.transformation.for.gehje}) the operator
$Q^2$
may be reduced to the form $\widetilde Q^2=2\p_u-\widetilde\delta x_a\p_a$,
where
$\widetilde\delta\in\{0;1\}$ (the operator $Q^1$ does not change in this case).
Let us substitute the coefficients of the operator~$\widetilde Q^2$
into the equation~(\ref{det.eqs.for.gehje.3}) and
integrate it with respect to~$F$: $F=e^{\widetilde\delta u}h(u_t)$, where
$h$ is a smooth function of the variable $u_t$.
If  $\widetilde\delta=1$, and $h$ is not fixed, then the algebra~$A^{\rm ext}$
is truly two-dimensional
(the second case of an extension).

If the algebra~$A^{\rm ext}$ is non-commutative, we can assume, that
$[Q^1,Q^2]=2Q^1$, whence~$\xi^0_t=2$, $\eta_t=0$, and by the
transformation~(\ref{partial.equivalence.transformation.for.gehje}) the
operator~$Q^2$
can be reduced to the form $\widetilde Q^2=2t\p_t+2u\p_u+\delta x_a\p_a$.
Let us substitute coefficients of the operator~$\widetilde Q^2$
to the equation~(\ref{det.eqs.for.gehje.3}) and
integrate it with respect to $F$: $F=|u|^{2-\delta}h(u_t)$, where
$h$ is a smooth function of the variable~$u_t$.
If  $\delta\not=2$, and $h$ is not specified, then the algebra $A^{\rm ext}$ is
truly two-dimensional
(the third case of an extension).

Let us assume that $\dim A^{\rm ext}=3,$ and
\[A^{\rm ext}=\langle  Q^i=\delta_i x_a\p_a+\xi^{0i}(t,u)\p_t+\eta^i(t,u)\p_u,
\; i=1,2,\; Q^3=\p_t\rangle, \]
the functions $\eta^1$ and $\eta^2$ are linearly independent,
and the equation~$\eta^1\cdot(\ref{det.eqs.for.gehje.3}.2)-
\eta^2\cdot(\ref{det.eqs.for.gehje.3}.1)$
is an identity with respect to~$F$ (here $(\ref{det.eqs.for.gehje.3}.i)$ is the
equation,
obtained from~(\ref{det.eqs.for.gehje.3}) by substitution of coefficients of the
operator $Q^i$).
By virtue of the latter condition the coefficients of the operators $Q^1$ and
$Q^2$ have to satisfy
the following equations:
\bss
\eta^1\eta^2_t=\eta^2\eta^1_t, \quad \eta^1(\eta^2_u-\delta_2)=\eta^2(\eta^1_u-
\delta_1), \\[1.2ex]
\eta^1(\eta^2_u-\xi^{02}_t)=\eta^2(\eta^1_u-\xi^{01}_t), \quad
\eta^1\xi^{02}_u=\eta^2\xi^{01}_u,
\ess
from which it follows that $(\delta_1,\delta_2)\not=(0,0)$, $\eta^i_t=0$,
$\xi^{0i}_t=\delta_i$, $i=1,2$,
$\eta^1=(\delta_1-\delta_2\varphi)/\varphi_u$, $\eta^2=\varphi\eta^1$ for a
certain
function~$\varphi=\varphi(u)\not=\const.$ Let us substitute, if necessary, the
operators $Q^1$ and $Q^2$ by their
linear combinations~$\widetilde Q^1$ and~$\widetilde Q^2$ so
$\widetilde \delta_1=1$, $\widetilde \delta_2=0$.
By the transformation~(\ref{partial.equivalence.transformation.for.gehje})
the operators~$\widetilde Q^1$ and~$\widetilde Q^2$
are reduced to the form $\widehat Q^1=t\p_t+u\p_u+\delta x_a\p_a$, $\widehat
Q^2=\p_u$.
After substitution of the coefficients of the operators $\widehat Q^1$
and~$\widehat Q^2$
into the equation~(\ref{det.eqs.for.gehje.3}) we obtain one equation for the
function $F$: $F_u=0$,
whence $F=h(u_t)$ where $h$ is a smooth function of the variable $u_t$.
If $h$ is not specified then the algebra $A^{\rm ext}$ is truly three-
dimensional (the fourth case of extension).

\vspace{1ex}

\noindent
{\bf Lemma 1.} In all other cases when $\dim A^{\rm ext}>1$, the function $F$
satisfies the equation
\be{eq.for.F}
(Au_t^2+Bu_t+C)F_{u_t}=(2Au_t+D)F,
\ee
where $A$, $B$, $C$ and $D$ are arbitrary smooth functions of the variable $u$.
The transformation~(\ref{partial.equivalence.transformation.for.gehje}) is an
equivalence
transformation at the set of equations of the form~(\ref{eq.for.F}).

\vspace{1ex}

When we integrate the equations~(\ref{eq.for.F}), four non-equivalent cases
arise where $F_{u_tu_tu_t}\not=0$.

\vspace{1ex}

1. $F=\alpha(u)e^{u_t}$  where $\alpha>0$. In this case an additional extension
of
the symmetry is possible only for $\alpha=A|u|^\nu e^{\mu u}$,
where $A,\nu,\mu=\const$, $A>0$.
By means of scale transformations by variables $x_a$ we can get $A=1$.
If \mbox{$\nu=0$}, then we can put $\mu=0$
(if it is not so, it is sufficient to perform the transformation
$\widetilde t=e^{\mu t}/\mu$, $\widetilde u=e^{\mu t}(u+2t-2/\mu)$;
here and in the following variables $x_a$ are not transformed if it is not
specifically mentioned).
Thus we obtain the fifth case of an extension.
For $\mu\nu\not=0$ an additional symmetry extension will arise only for $\nu=2$,
but then it is possible to put $\mu=0$ (the respective equivalence
transformation is
$\widetilde t=e^{\mu t}/\mu$, $\widetilde u=e^{\mu t}u$).
If $\nu\not=0$ and $\mu=0$, then the equation~(\ref{gehje}) has the same
symmetry
algebra as in the more general third case of an extension.

\vspace{1ex}

2. $F=\alpha(u)|u_t+\delta|^{\beta(u)},$  where $\delta\in\{0;1\}$, $\alpha>0$.
For additional symmetry extension it is necessary $\beta=\const$
(and $\beta\not\in\{0;1;2\}$, or otherwise $F_{u_tu_tu_t}=0$).
Then any equation~(\ref{gehje}) with  $\delta=0$ is equivalent to the same
equation in which in
addition $\alpha=1$ (the sixth case of an extension).
The equation~(\ref{gehje}) with $\delta\not=0$ is also reduced to the same case
of extension, if
$(\alpha'\alpha)'=0$. Really, then $\alpha=Ae^{\mu u}$ and we can put $A=1$
(due to scal transformations by the variables $x_a$),
$\mu=0$ and $\delta=0$ (the respective equivalence transformation when
$\mu\not=0$ has the form
$\widetilde t=e^{\mu\delta t/\beta}\beta/(\mu\delta)$,
$\widetilde u=e^{\mu(u+\delta t)/(\beta-2)}(\beta-2)/\mu$,
and when $\mu=0$ it has the form $\widetilde t=t$, $\widetilde u=u+\delta t$).
If $\delta\not=0$ and $(\alpha'\alpha)'\not=0$, then we will not obtain new
cases of symmetry extension.

\vspace{1ex}

3. $F=\alpha(u)e^{\beta(u)\tan^{-1} u_t},$  where $\alpha>0$,
$\beta\not\equiv0$.
An additional extension of the symmetry exists only on the conditions
$\beta=\const$ and $\alpha$ being an exponential or a power function,
but it is not wider than extensions for more general cases 2--4 from the
Table~1, and for this reason
there is not need to list this case separately.

\vspace{1ex}

4. $F=\alpha(u)u_t^2e^{u_t^{-1}},$  where $\alpha>0$.
The equation with such second part is equivalent to the equation with
$F=e^{u_t}$.
The equivalence is determined by the transformation
(of the type~(\ref{general.equivalence.transformation.for.gehje}))
$\widetilde t=u$, $\widetilde u=t+\int\ln\alpha(u)du$.

\vspace{1ex}

We completed consideration of the case $F_{u_tu_tu_t}\not=0$.

\vspace{1ex}

Let in the following {\mathversion{bold}$F_{u_tu_tu_t}=0$}, whence
\be{quadric.F}F=A(t,u)u_t^2+B(t,u)u_t+C(t,u),\ee
where $A$, $B$, $C$ are smooth functions of the variables $t$ and $u$.
Splitting the equations~(\ref{det.eqs.for.gehje.2})
and~(\ref{det.eqs.for.gehje.3})
by the variable~$u_t$, in addition to~(\ref{det.eqs.for.gehje.1}) we obtain the
following system
of determining equations:
\be{det.eqs.for.quadric.gehje.2}
\xi^0_a=A\xi^a_t-\tfrac{1}{2}B\xi^a_u, \quad \eta_a=C\xi^a_u-
\tfrac{1}{2}B\xi^a_t,
\ee
\bs{det.eqs.for.quadric.gehje.3}
A_t\xi^0+A_u\eta+B\xi^0_u=2A(\xi^0_t-\xi^1_1), \\[1.2ex]
C_t\xi^0+C_u\eta+B\eta_t=2C(\eta_u-\xi^1_1), \\[1.2ex]
B_t\xi^0+B_u\eta+2A\eta_t+2C\xi^0_u=B(\eta_u+\xi^0_t-2\xi^1_1).
\es

The equivalence group of the set of equations~(\ref{gehje}) with second parts
quadratic by $u_t$
coincides with the general equivalence group of the equations~(\ref{gehje}). By
means of
the transformation~(\ref{general.equivalence.transformation.for.gehje})
the coefficients $A$, $B$ and $C$ are changed as follows:

\bs{ABC.transformation}
\delta^2\widetilde A=A\zeta_t^2-B\zeta_t\zeta_u+C\zeta_u^2, \\[1.2ex]
\delta^2\widetilde B=B(\zeta_t\varphi_u+\zeta_u\varphi_t)-2A\zeta_t\varphi_t-
2C\zeta_u\varphi_u, \\[1.2ex]
\delta^2\widetilde C=A\varphi_t^2-B\varphi_t\varphi_u+C\varphi_u^2,
\es

The condition $B^2-4AC=0$ is invariant with respect to transformations of the
type~(\ref{ABC.transformation}),
that are generated by equivalence
transformations~(\ref{general.equivalence.transformation.for.gehje}),
and for this reason it is a classifying condition.
If $B^2-4AC=0$, the we can put $A=1$, $B=0$, whence $C=0$
the seventh case of an extension.
In the following $B^2-4AC\not=0$.

\vspace{1ex}

\noindent
{\bf Lemma 2.} On the condition $B^2-4AC\not=0$ it follows from the
equations~(\ref{det.eqs.for.gehje.1})
and~(\ref{det.eqs.for.quadric.gehje.2})
that coefficients of any symmetry operator of the equation~(\ref{gehje}) have
the following form:
\bs{operator.coefficients.for.quadric.gehje}
\xi^a=2\gamma_bx_bx_a-\gamma_ax_bx_b+\sigma_{ab}x_b+\beta x_a+\alpha^a,
\\[1.2ex]
\xi^0=\tfrac{1}{2}(A\beta_t-\tfrac{1}{2}B\beta_u)x_ax_a+(A\alpha^a_t-
\tfrac{1}{2}B\alpha^a_u)x_a+\alpha^0,
\\[1.2ex]
\eta=\tfrac{1}{2}(C\beta_u-\tfrac{1}{2}B\beta_t)x_ax_a+(C\alpha^a_u-
\tfrac{1}{2}B\alpha^a_t)x_a+\alpha^4,
\es
where $\gamma_a$, $\sigma_{ab}$ are constants,
$\beta$, $\alpha^0$, $\alpha^a$, $\alpha^4$ are smooth functions of the
variables $t$ and $u$.

\vspace{1ex}

Let us substitute the
expressions~(\ref{operator.coefficients.for.quadric.gehje}) into the
system~(\ref{det.eqs.for.quadric.gehje.3}) and split it by the variables $x_a$.
As a result we obtain $n+2$ systems of the same structure
\bs{det.eqs.for.quadric.gehje.3a}
H^1A_t+H^2A_u+H^1_uB=2(H^1_t-\lambda)A,\\[1.2ex]
H^1C_t+H^2C_u+H^2_tB=2(H^2_u-\lambda)C,\\[1.2ex]
H^1B_t+H^2B_u+2H^2_tA+2H^1_uC=(H^1_t+H^2_u-2\lambda)B,
\es
where $H^1$, $H^2$ and $\lambda$ take the following values:
\be{det.eqs.for.quadric.gehje.3aH} \arraycolsep=0em\begin{array}{lll}
H^1=A\beta_t-\tfrac{1}{2}B\beta_u, & \qquad H^2=C\beta_u-\tfrac{1}{2}B\beta_t, &
\qquad \lambda=0; \\[1.2ex]
H^1=A\alpha^a_t-\tfrac{1}{2}B\alpha^a_u, & \qquad H^2=C\alpha^a_u-
\tfrac{1}{2}B\alpha^a_t, &
\qquad \lambda=2\gamma_a; \\[1.2ex]
H^1=\alpha^0, & \qquad H^2=\alpha^4, & \qquad \lambda=\beta.
\end{array}\ee
Investigating the systems of determining
equations~(\ref{det.eqs.for.quadric.gehje.3a}),
(\ref{det.eqs.for.quadric.gehje.3aH}),
we obtain the following statements.

\vspace{1ex}

\noindent
{\bf Lemma 3.} If $\dim A^{\rm max}>\dim A^{\rm ker}$, then the
equation~(\ref{gehje}), (\ref{quadric.F})
is equivalent to the same equation in which in addition $A=e^{\delta t}\widehat
A(u)$,
$B=e^{\delta t}\widehat B(u)$ and $C=e^{\delta t}\widehat C(u)$,
and for which $A^{\rm max}\ni2\p_t-\delta x_a\p_a$
(with functions $\widehat A(u)$, $\widehat B(u)$, $\widehat C(u)$ being not
specified,
$A^{\rm max}=\langle\p_a,\: J_{ab},\: 2\p_t-\delta x_a\p_a\rangle$, and it is
the first case of an extension).

\vspace{1ex}

\noindent
{\bf Lemma 4.} If $\dim A^{\rm max}>\dim A^{\rm ker}+1$, then the
equation~(\ref{gehje}), (\ref{quadric.F})
is equivalent to the same equation
in which in addition $A_t=0$, $B_t=0$ and $C_t=0$.

\vspace{1ex}

It follows from the Lemmas 3 and 4 that for completion of the classification it
is necessary to
investigate the equation~(\ref{gehje}) with the second parts of the form
$F=A(u)u_t^2+B(u)u_t+C(u)$,
where $A$, $B$ and $C$ are smooth functions of the variable $u$, $B^2-
4AC\not=0$.
The equivalence group of the set of these equations coincides with the general
equivalence
group of the equations~(\ref{gehje}), whose second parts do not depend on $t$.
Under the
action of the
transformation~(\ref{partial.equivalence.transformation.for.gehje})
the coefficients $A$, $B$ and $C$ change as follows:
\bss
\delta^2\widetilde A=A\hat\delta^2-B\hat\delta \zeta_u+C\zeta_u^2, \quad
\delta^2\widetilde B=B\hat\delta \varphi_u-2C\zeta_u\varphi_u, \quad
\delta^2\widetilde C=C\varphi_u^2,
\ess
and we can assume additionally that $B=0$, $C=\varepsilon_1=\pm 1$.
Thus in the following $F=A(u)u_t^2+\varepsilon_1$ where $A\not=0$.

The condition $A_u=0$ gives the eighth case of a symmetry extension.

The case when  $\exists\mu=\const$: $(u+\mu)A_u+2A=0$,
is reduced to the case $A_u=0$ by means of the equivalence transformation
$\widetilde t=(u+\mu)\sinh t$, $\widetilde u=(u+\mu)\cosh t$, if
$\varepsilon_1A<0$,
or $\widetilde t=(u+\mu)\sin t$, $\widetilde u=(u+\mu)\cos t$, if
$\varepsilon_1A>0$.

The case when $\exists\mu,\nu=\const$ $(\nu\not=0,2)$: $(u+\mu)A_u+\nu A=0$,
is reduced to a more general case~3 from the Table~1 by means of the equivalence
transformation
$\widetilde t=t$, $\widetilde u=|u+\mu|^{1-\nu/2}$.

The condition $(A_u/A)_u=0$, $A_u\not=0$ gives the ninth case of a symmetry
extension.

Let the function $A$ satisfy the equation $A_u/A=\nu A+\mu$ for some non-zero
constants
$\mu$ and $\nu$. Any solution of this equation depending on values of the
constants $\mu$ and $\nu$
is equivalent (by translations and scale transformations by the variable $u$)
to one of the following functions:
\be{trigonom.and.hyperbol.right.part}
\frac{\varepsilon_2}{\cosh^2 u}, \quad
\frac{\varepsilon_2}{\sinh^2 u}, \quad
\frac{\varepsilon_2}{\cos^2  u}, \quad
\ee
where $\varepsilon_2=\pm 1,$ $(\varepsilon_1,\varepsilon_2)\not=(-1,-1)$.
Let $\varepsilon_0=1$ for the first two functions, and
$\varepsilon_0=-1$ for the third function.

\vspace{1ex}

\noindent
{\bf Lemma 5.} The equation~(\ref{gehje}) with the second parts
$A(u)u_t^2+\varepsilon_1$ and $\widetilde A(u)u_t^2+\widetilde\varepsilon_1$,
where $A$ and $\widetilde A$ are chosen from the set of
functions~(\ref{trigonom.and.hyperbol.right.part}),
are equivalent if and only if
$\varepsilon_0\widetilde\varepsilon_0=\varepsilon_1\widetilde\varepsilon_1=
\varepsilon_2\widetilde\varepsilon_2$.

\vspace{1ex}

By virtue of Lemma~5 the cases~10--12 of the Table~1 exhaust all possible non-
equivalent equations from this class.

For any other functions $A=A(u)$ there would be no extension of the symmetry for
the equation~(\ref{gehje}).

Non-equivalence of the all cases of extensions adduced in the Table~1,
where it was not proved directly by means of application of equivalence
transformations, follows explicitly
from non-isomorphness of the respective maximal symmetry algebrae, in
particular, from the fact
they have different dimensions.

\vspace{1ex}

\noindent
{\bf 7. Conclusion.}
We presented a complete solution of group classification problem for a class of
PDEs with
derivatives entering nonlinearly.
New invariant equations obtained in this paper are very interesting also from
the point of view
 of finding new conditional symmetries~\cite{fshs.eng} for higher order
equations,
if these new equations are used as differential constraints. In particular, this
method
works for nonlinear wave equations, and that will be the subject of further
papers.


\renewcommand{\refname}{\vspace{-5ex}\par}


\begin{thebibliography}{99}

\bibitem{ovsiannikov.eng}
Ovsjannikov L V 1982 {\it Group Analysis of Differential Equations} (New York:
Academic Press)

\bibitem{lie} Lie S 1891 {\it Vorlesungen \"uber Differentialgleichungen mit
Bekannten Infinitesimalen
Transformationen} (Leipzig: B.G. Teubner)

\bibitem{akhatov&gazizov&ibragimov.dan.1987.eng}
Akhatov I S, Gazizov R K and Ibragimov N K 1987 {\it Proc. Acad. Sci. USSR} {\bf
293} 1033--5

\bibitem{dorodnitsyn}
Dorodnitsyn V A 1982 {\it Zhum. Vych. Maternal Matem. Fiziki} {\bf 22} 1393--400
\bibitem{oron&rosenau}
Oron A and Rosenau P 1986 {\it Phys. Lett. A} {\bf 118} 172-6
\bibitem{edwards}
Edwards M P 1994 {\it Phys. Lett. A} {\bf 190} 149-54
\bibitem{gandarias.jphysa.96}
Gandarias M L 1996 {\it J. Phys. A: Math. Gen.} {\bf 29} 607--33

\bibitem{olver&heredero}
Olver P J and Heredero R H 1996 {\it J. Math. Phys.} {\bf 37} 6419--38

\bibitem{ibragimovtorrisi.jmp.91}
Ibragimov N K, Torrisi M and Valenti A 1991 {\it J. Math. Phys.} {\bf 32} 2988--
95
\bibitem{ibragimovtorrisi.jmp.92}
Ibragimov N K and Torrisi M 1992 {\it J. Math. Phys.} {\bf 33} 3931--7
\bibitem{torrisi.jmp.96}
Torrisi M, Tracina R and Valenti A 1996 {\it J. Math. Phys.} {\bf 37} 4758--67
\bibitem{torrisi.jnm.98}
Torrisi M and Tracina R 1998 {\it Int. J. Nonlinear Mech.}  {\bf 33} 473--87

\bibitem{serov&cherniha.umj.97}
Serov M I and Cherniha R M 1997 {\it Ukrain. Math. J.} 49 1262--70

\bibitem{nikitin&wiltshire.conf99}
Nikitin A G and Wiltshire R J 2000 {\it Proc. of Institute of Mathematics of NAS
of Ukraine} vol~30 part~1
(Kyiv: Institute of Mathematics of NAS of Ukraine) 47--59.


\bibitem{akhatov&gazizov&ibragimov.eng}
Akhatov I S, Gazizov R K and Ibragimov N K 1989 {\it Sovremennye Problemy
Matematiki. Novejshie Dostizheniya}
 vol 34 (Moscow: Nauka) pp 3--83

\bibitem{zhdanov&lahno.jphys.a.1999}
Zhdanov R Z and Lahno V I 1999 {\it J.~Phys.~A: Math.~Gen.} {\bf 32} 7405--18

\bibitem{zhdanov&roman.rep.math.phys.2000}
Zhdanov R Z and Roman O V 2000 {\it Rep. Math. Phys.} {\bf 45} no~2 273--91

\bibitem{zb2000.with.cherniha}
Popovych R and Cherniha R 2001 {\it Proc. of Institute of mathematics of NAS of
Ukraine}
(Kyiv: Institute of Mathematics of NAS of Ukraine), to be published

\bibitem{boyer&penafiel}
Boyer C P and Penafiel M N 1976 {\it Nuovo cim.~B} {\bf 31}, N~1 195--210

\bibitem{fushchich&shtelen.lett.nuovo.cim.1982}
Fushchich W I and Shtelen W M 1982 {\it Lett. nuovo cim.} {\bf 34}, N~16 498.

\bibitem{fshs.eng}
Fushchych W I, Shtelen W M and Serov N I 1989 {\it Symmetry Analysis and Exact
Solutions of Nonlinear Equations
 of Mathematical Physics} (Kiev: Naukova Dumka)

Fushchych W I, Shtelen W M and Serov N I 1992 {\it Symmetry Analysis and Exact
Solutions of Nonlinear Equations
 of Mathematical Physics} (Dordrecht: Kluwer Academic Publisher) (English
transl.)

\bibitem{fs.danu.90.7}
Fushchych W I and Serov N I 1991 {\it Ukrain. Math. J.} 43 394--399

\bibitem{olver.eng}
Olver P J 1986 {\it Applications of Lie Groups to Differential Equations}
(Berlin: Springer-Verlag)

\end{thebibliography}
\end{document}